\documentclass[10pt,aps,prd,twocolumn,amsmath,amssymb,nofootinbib,longbibliography]{revtex4-1}
\usepackage{amsfonts,amsthm,mathrsfs}
\usepackage{lipsum,graphicx}
\usepackage[usenames,dvipsnames,svgnames]{xcolor}
\usepackage{dutchcal}
\usepackage{times}
\usepackage{inputenc}
\usepackage{bm}
\usepackage[normalem]{ulem}
\usepackage{multirow}
\usepackage{url}
\usepackage{natbib}
\usepackage{subfigure}
\usepackage[colorlinks=true,citecolor=RoyalBlue,urlcolor=RoyalBlue,linkcolor=RoyalBlue]{hyperref}
\begin{document}
\preprint{}
\title{Thermal dilepton production within conformal viscous Gubser flow}

\author{Lakshmi J. Naik}
\email{jn\_lakshmi@cb.students.amrita.edu}

\author{V. Sreekanth}
\email{v\_sreekanth@cb.amrita.edu}


\affiliation{Department of Physics, Amrita School of Physical Sciences, Coimbatore, Amrita Vishwa Vidyapeetham, India}

\date{\today}
\begin{abstract}
 By employing the Gubser solutions of causal relativistic second-order Israel-Stewart hydrodynamics, we study the thermal dilepton production from heavy-ion collisions, considering the transverse expansion of the viscous hot QCD medium along with longitudinal boost-invariance. We analyze the evolution of the temperature and shear stress profiles of the QCD matter under Gubser flow for different values of the associated parameter $q$ (inverse length scale). We study the dilepton production using leading order Born rates from QGP and hadronic sectors under Gubser geometry. Viscous modified dilepton rate is calculated using the first-order Chapman-Enskog (CE) like non-equilibrium correction of the particle distribution function. 
Our study indicates that lower values of $q$ result in the enhancement of the emitted dilepton spectra. 
We also determine the effective temperature of the hot QCD medium from the inverse slope of transverse mass spectra, for different $q$. We find that the effective temperature determined from the dilepton spectra for a smaller system to be higher. Further, we compare the strength of CE like and Grad's viscous correction to the ideal dilepton spectra and find that CE type viscous corrections 
are well behaved compared to that of Grad's in the presence of transverse flow.
\end{abstract}
\maketitle

\section{Introduction}

The ultra-relativistic heavy-ion collision experiments provide an opportunity to explore various aspects of the strongly interacting hot and/or dense nuclear matter called Quark-Gluon Plasma (QGP). Analysis of the experimental data from these collisions imply that the QGP behaves like a near prefect fluid with extremely low value of shear viscosity to entropy density ratio, $\eta/s = 1/4\pi$~~\cite{STAR:2005gfr,PHENIX:2004vcz,BRAHMS:2004adc,PHOBOS:2004zne,Kovtun:2004de,Hirano:2005wx,Romatschke:2007mq,ALICE:2010suc,Heinz:2013th}. This has invoked interest in employing relativistic dissipative hydrodynamics to model the expansion of QGP fireball in heavy ion collisions~\cite{Busza:2018rrf,Romatschke:2017ejr}. Hydrodynamics describes the QGP very close to thermal equilibrium and QCD kinetic theory calculations bring an insight on how the system attains this 
thermal equilibrium~\cite{Berges:2020fwq}.

It has been now well accepted that the evolution of QGP can be described using relativistic viscous hydrodynamics~\cite{Luzum:2008cw,Jaiswal:2016hex,Rocha:2023ilf}. The relativistic first-order Navier-Stokes theory~\cite{Landau:1987,Eckart:1940te} exhibits acausal behaviour~\cite{Hiscock:1983zz,Hiscock:1985zz,Muronga:2001zk} and this has led to the development of causal higher order dissipative theories. Recently, several second-order relativistic dissipative hydrodynamic theories have been formulated and applied in the context of heavy-ion collisions~\cite{Rocha:2023ilf}.
Some of the relativistic second-order dissipative hydrodynamic formalisms derived  can be found in following works~\cite{Muller1967329,Israel:1976tn,Israel:1979wp,Muronga:2003ta,Denicol:2010xn,Denicol:2012cn,Baier:2007ix,Jaiswal:2012qm,Jaiswal:2013fc}. Similarly, several formalisms of relativistic third-order theories based on different frameworks also exist~\cite{El:2009vj,Younus:2019rrt,deBrito:2023tgb,Panday:2024hqp}. Recently, there have been attempts to develop first-order causal viscous theories as well~\cite{Van:2011yn,Bemfica:2017wps,Kovtun:2019hdm,Biswas:2022cla}.

The relativistic hydrodynamic equations can be solved by considering the geometry of heavy-ion collisions~\cite{Belenkij:1955pgn,Landau:1953gs}. The expansion of QGP is initially along the collision axis ($z$-axis) and later, transverse expansion also starts to build up. The simplest model which describes the dynamics of heavy-ion collisions at early proper times is the one-dimensional Bj\"orken flow~\cite{Bjorken:1982qr}. The Bj\"orken's solution possess boost invariance along the longitudinal direction ($z$-axis) and translational invariance in the transverse plane. Gubser flow is a two-dimensional model considering both the longitudinal and transverse expansion of QGP, developed in Refs.~\cite{Gubser:2010ze,Gubser:2010ui}. This model provides semi-realistic solutions to the hydrodynamic equations by invoking the conformal equation of state ($\epsilon = 3P$). The Gubser solutions are symmetric under the $SO(3)_q \otimes S(1,1) \otimes \textbf{Z}_2$ group transformations. 

Gubser's solution has been employed in the context of heavy-ion collisions to study various aspects of the matter produced. Solutions of relativistic ideal hydrodynamics and first-order Navier-Stokes equation within the Gubser flow have been found and analysed ~\cite{Gubser:2010ze,Gubser:2010ui}. In Ref.~\cite{Marrochio:2013wla}, solutions of second-order Israel-Stewart (IS) theory have been derived within the same model. Later, an exact solution of relativistic Boltzmann equation was studied in the relaxation time approximation for system evolving under Gubser symmetry and it was found that the second-order hydrodynamic theories provide an overall good agreement with this solution~\cite{Denicol:2014tha,Denicol:2014xca}. Further, solutions of anisotropic hydrodynamic equations have also been explored considering the Gubser flow in several works~\cite{Nopoush:2014qba,Martinez:2017ibh,Chattopadhyay:2018apf}. Within the same model, the analytical solutions of dissipative spin hydrodynamics have also been looked at recently~\cite{Wang:2021wqq}.

A completely analytical description of elliptic flow $(v_2)$ in relativistic dissipative hydrodynamics within Gubser model was obtained in Ref.~\cite{Hatta:2014upa}. Effect of magnetic field on directed flow $v_1$ was also investigated within Gubser flow~\cite{Gursoy:2014aka}. 
In Ref.~\cite{Hatta:2014jva}, the flow harmonics have been calculated analytically within an anisotropically perturbed Gubser flow and later, in Ref.~\cite{Hatta:2015era}, they were derived at finite density.
Gubser flow has been employed in the study of quarkonia suppression in small systems~\cite{Bagchi:2023vfv}.
Recently, thermal photon production within Gubser flow was analysed in Ref.~\cite{Paquet:2023bdx}, where the effect of viscosity was not included. In the present work, we intend to study another important signal from heavy-ion collisions - thermal dileptons, 
by considering the viscous QGP expansion modelled using Gubser flow within causal second-order IS relativistic hydrodynamics.

Electromagnetic signals such as dileptons and photons are emitted from all the stages of heavy-ion collisions. Since they escape from the fireball easily, they remain an important probe for the 
analysis of the properties of high temperature matter created~\cite{Rapp:2011is,Vujanovic:2016anq,David:2019wpt,Salabura:2020tou,Gale:2021emg,Geurts:2022xmk,Coquet:2021gms,Coquet:2023wjk}. Photon and dilepton production from different stages of the QGP evolution is studied by identifying major contributory mechanisms~\cite{Baier:1988xv,Haglin:1992fy,Aurenche:1998nw,Arnold:2001ms,Arnold:2001ba,Arnold:2002zm,Ghiglieri:2014kma,Heffernan:2014mla,Gotz:2021dco,Churchill:2023vpt}. They can emanate majorly from the initial hard scatterings, medium induced thermal radiation and final stage hadronic decays and have proved to be a great source in understanding the properties of matter created~\cite{Rapp:2011is,David:2019wpt,Geurts:2022xmk}. Recently, photons and dileptons are shown to be 
very useful to probe the early-time pre-equilibrium states of the fireball as well~\cite{Oliva:2017pri,Coquet:2021lca,Churchill:2020uvk,Wu:2024igf,Garcia-Montero:2023lrd,Garcia-Montero:2024lbl}.

\par 
Dileptons, despite being produced less compared to the photons, are advantageous as they have invariant mass ($M$) and hence can be separated from various sources. The presence of invariant mass leaves their spectra unaffected by the Doppler shifts, unlike the case of photons~\cite{Rapp:2014hha}. Drell-Yan process in the early times produces dileptons having very large invariant mass~\cite{Drell:1970wh}. Dileptons from the QGP phase are dominantly produced in the intermediate mass range $1.1 \leq M (\textrm{GeV}) \leq 2.9$, while those from the hadronic decays are created in the low mass range $M<1.1$ GeV~\cite{Rapp:2011is,Salabura:2020tou,Geurts:2022xmk}.
In the QGP phase, the most dominant contribution is known to come from the thermal dileptons produced via the $q\bar{q}$-annihilation. Another major source for thermal dileptons comes from the pion annihilation in the hadronic phase. 
The presence of dissipation in the QGP medium affects the production of these particles. 
The impact of viscosity on thermal particle production has been investigated using causal relativistic hydrodynamics and found to show appreciable effects on the spectra~\cite{Dusling:2008xj,Bhatt:2009zg,Bhatt:2010cy,Bhatt:2011kx,Bhalerao:2013aha,Vujanovic:2013jpa,Chandra:2015rdz,Naik:2022pyk}.

In the present work, we study the thermal dilepton spectra in presence of shear viscosity from heavy-ion collisions using the solutions of second-order IS theory within Gubser model. The effect of viscosity on the dilepton production rate is incorporated through the Chapman-Enskog like non-equilibrium distribution function which is known to have a rapid convergence up to second order~\cite{Bhalerao:2013pza}. Further, we determine an effective temperature of the QGP medium from analysing the obtained dilepton yield. 

The paper is structured as follows. In Section~\ref{Sec:Gubser_flow}, we review the Gubser model. 
Causal second-order dissipative hydrodynamics of Israel-Stewart within Gubser flow is presented in Section~\ref{Sec:Israel-Stewart}. 
We evaluate the dilepton yield within Gubser flow in Section~\ref{Sec:dilepton-spectra}. Section~\ref{Sec:results} is devoted to the results and its discussion. We summarize our results in Section~\ref{Sec:conclusion}.

{\it Notations and conventions:} Throughout the manuscript, we follow the metric convention $g_{\mu\nu} = \textrm{diag}(+1,-1,-1,-1)$. We take $c=1$ in our calculations.

\section{Gubser flow} \label{Sec:Gubser_flow}

Milne coordinates $x^\mu = (\tau, r, \phi, \eta_s)$ are the most natural choice to describe the heavy-ion collision scenario at ultra-relativistic energies and are expressed as
\begin{eqnarray}
    \tau &=& \sqrt{t^2 - z^2}, \quad \eta_s = \tanh^{-1}(z/t), \nonumber\\
    r &=& \sqrt{x^2 + y^2}, \quad\, \phi = \tan^{-1}(y/x).
\end{eqnarray}
Here, $\tau$ and $\eta_s$ denote the proper time and space-time rapidity respectively, $r$ is the radial distance from the fireball center and $\phi$ is the azimuthal angle in the transverse plane. The Bj\"orken flow model~\cite{Bjorken:1982qr} offers a simplified, yet essential, one-dimensional representation of QGP dynamics and evolution. This model
can be expressed in terms of the above coordinate system with the line element given by $ds^2 = d\tau^2 -dr^2 - r^2 d\phi^2 - \tau^2 d\eta_s^2$ and fluid four-velocity $u^\mu=(1,0,0,0)$.  
The Bj\"orken's prescription is based on several assumptions such as boost invariance along $\eta_s$ direction, translational invariance in the transverse plane and symmetry under reflections along the $\eta_s$ direction. Steven Gubser made a generalization of Bj\"orken flow by replacing the translational invariance with the conformal symmetry $SO(3)$, while the invariance under boosts and reflections still maintained~\cite{Gubser:2010ze}. Under the Gubser model, the system has a finite transverse size and it undergoes expansion along both radial and longitudinal directions. However, such a description is applicable only for a system of conformal fluids. 

Gubser flow on a de-Sitter background is obtained by Weyl rescaling of the metric measure in Milne coordinates, $i.e.,$ $ds^2 \rightarrow d\hat{s}^2 \equiv ds^2/\tau^2$, which is followed by a coordinate transformation from $(\tau, r)$ to the {\it Gubser coordinates} $(\rho,\theta)$:
\begin{eqnarray}
    \sinh\rho &\equiv& -\frac{1-(q\tau)^2+(qr)^2}{2q\tau}, \label{Eq:rho}\\
    \tan\theta &\equiv& \frac{2qr}{1+(q\tau)^2-(qr)^2}. \label{Eq:theta}
\end{eqnarray}
Note that, here $q$ describes an arbitrary energy scale which 
represent the transverse size of the system. Note that, in the following, all the variables depending on the Gubser coordinates are denoted by a hat. 
Now, the Weyl rescaled line element in the new coordinates $\hat{x}^\mu = (\rho, \theta, \phi, \eta_s)$ is given by 
\begin{eqnarray}
  d\hat{s}^2 &=& d\rho^2 - (\cosh^2 \rho\,d\theta^2 + \cosh^2 \rho \sin^2\theta\,d\phi^2 + d\eta_s^2 ). \nonumber
\end{eqnarray}
In the de Sitter coordinates, Gubser flow appears to be static $i.e.,$ $\hat{u}^\mu = (1,0,0,0)$; whereas, 
the Weyl rescaling makes all the macroscopic quantities dimensionless by scaling with appropriate powers of rescaling parameter $\tau$. Further, the rescaling renders the fluid homogeneous, with all the fields depending only on the coordinate $\rho$. Note that the scalar expansion rate of the medium can be obtained as $\hat{\Theta} \equiv \hat{D}_\mu \hat{u}^\mu = 2\tanh\rho$. Below, we note the transformation relations followed by the hydrodynamic fields in our analysis:
\begin{eqnarray}
    T(\tau, r) &=& \frac{\hat{T}}{\tau}, \\
    \pi_{\mu\nu}(\tau, r) &=& \hat{\pi}_{\alpha\beta} \frac{1}{\tau^2} \frac{\partial \hat{x}^\alpha}{\partial x^\mu} \frac{\partial \hat{x}^\beta}{\partial x^\nu}.
\end{eqnarray}
%

\section{Israel-Stewart theory within Gubser flow} \label{Sec:Israel-Stewart}

\begin{figure*} 
  \centering
  \subfigure
  []{\includegraphics[width=8.5cm]{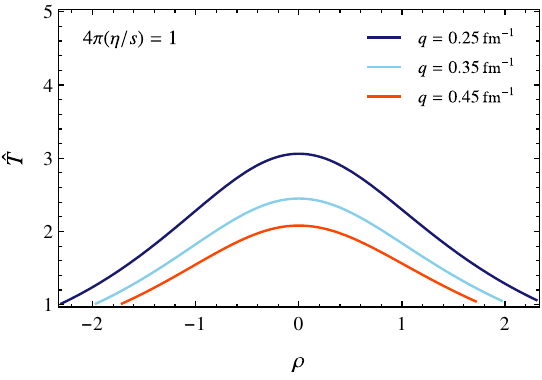}\label{fig:T-rho}} \quad
  \subfigure[]{\includegraphics[width=8.5cm,height=5.8cm]{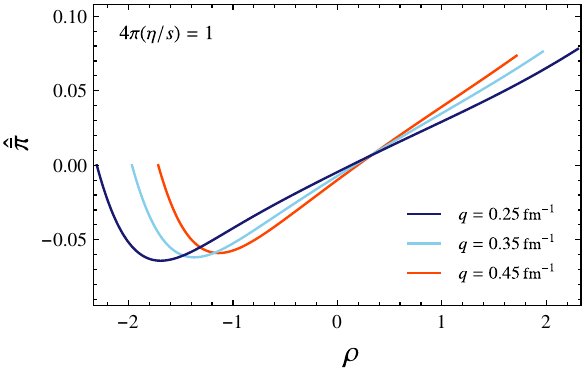}\label{fig:pi-rho} 
  } 
  \caption{(a) Temperature $\hat{T}$ and (b) normalized shear stress profile $\hat{\bar{\pi}} \equiv \hat{\pi}/(\hat{\epsilon} + \hat{P})$ of the hot QCD matter as a function of Gubser coordinate $\rho$ for different values of $q$, with $4\pi(\eta/s)=1$.}
  \label{fig:evolution-rho}
  \end{figure*}
\begin{figure*} 
  \centering
  \subfigure
  []{\includegraphics[width=8.5cm]{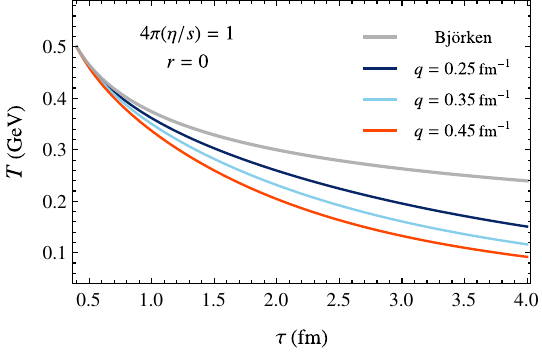}\label{fig:T-tau}} \quad
  \subfigure[]{\includegraphics[width=8.5cm,height=5.55cm]{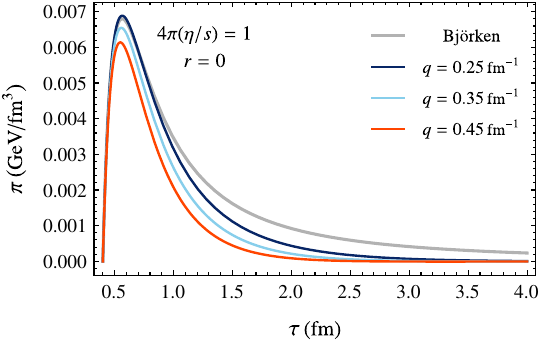}\label{fig:pi-tau} 
  } 
  \caption{(a) Temperature $T$ and (b) shear stress $\pi$ evolution of the hot QCD matter (at $r=0$) as a function of Milne coordinate $\tau$ (proper time) for different values of $q$, with $4\pi(\eta/s)=1$.}
  \label{fig:evolution-tau}
  \end{figure*}
\begin{figure}
    \centering
    \includegraphics[height=12.9cm]{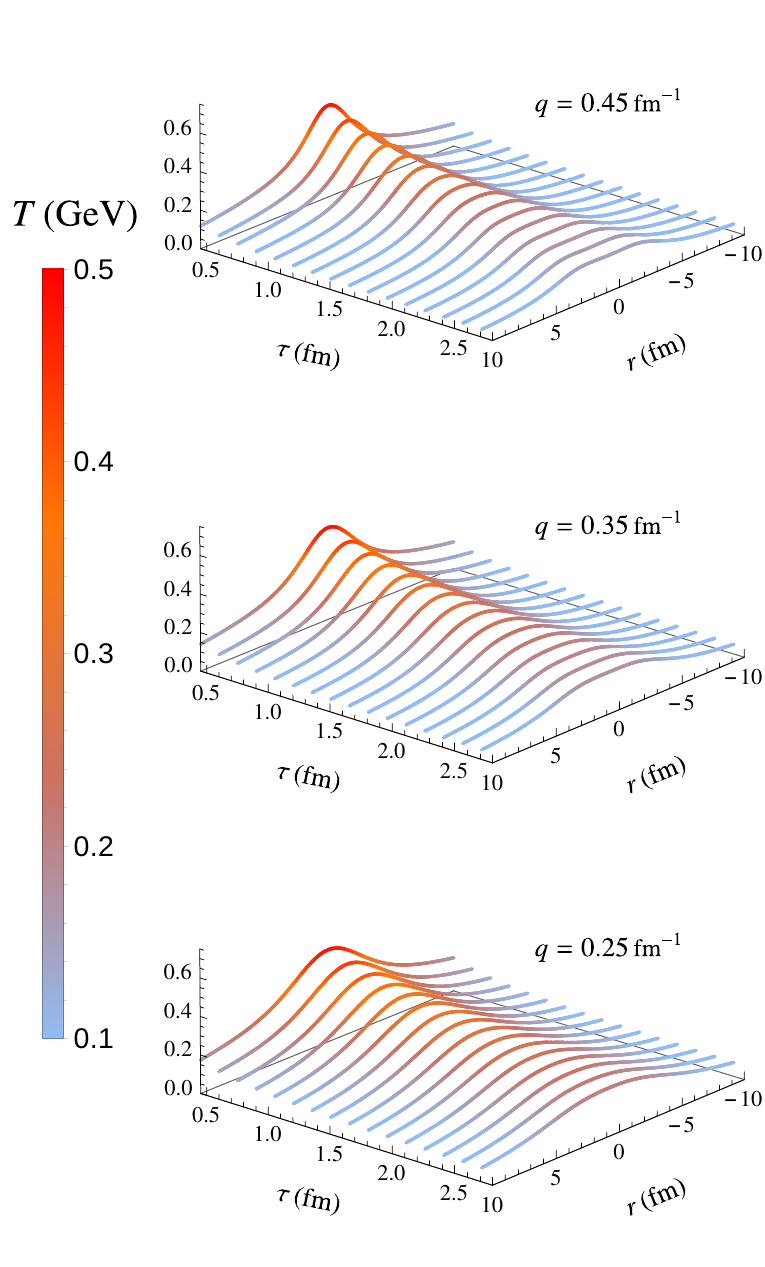}
    \caption{Temperature of QGP obtained by solving the IS hydrodynamics within Gubser flow, for different values of $q$. We have taken the viscosity to be $4\pi(\eta/s)=1$ for this analysis.}
    \label{fig:T-profile}
\end{figure}

In this section, we briefly review the second-order Israel-Stewart hydrodynamics formulated within the Gubser model as derived in Ref.~\cite{Marrochio:2013wla}. We begin by noting the energy-momentum tensor in Gubser coordinates for a viscous fluid with energy density $\hat{\epsilon}(\rho)$ and pressure $\hat{P}(\rho)$:
\begin{eqnarray}
    \hat{T}^{\mu\nu} = (\hat{\epsilon} + \hat{P})\hat{u}^\mu \hat{u}^\nu - \hat{P} \hat{g}^{\mu\nu} + \hat{\pi}^{\mu\nu},
\end{eqnarray}
where $\hat{\pi}^{\mu\nu}(\rho)$ is the shear-stress tensor. Recall that, because of Weyl rescaling, all the hydrodynamic quantities are dimensionless with functional dependence only on the coordinate $\rho$. 
Within this model, $\hat{\pi}_{\nu}^{\mu}$ is diagonal and is parameterized as
\begin{equation}
    \hat{\pi}_{\nu}^{\mu} = \textrm{diag}(0, \hat{\pi}/2, \hat{\pi}/2, -\hat{\pi}),
\end{equation}
We note that the form of shear stress tensor appears identical to that defined within Bj\"orken model. Also, $\hat{\pi}_\rho^\alpha = 0$ (for $\alpha = \rho,\,\theta,\,\phi,\,\eta_s$), which implies orthogonality to the flow profile. 
The evolution for $\hat{\epsilon}$ and $\hat{\pi}$ can be obtained by solving the nonlinear equations of motion~\cite{Marrochio:2013wla}:
\begin{eqnarray}
    \frac{d\hat{\epsilon}}{d\rho} + \frac{8}{3}\hat{\epsilon} \tanh\rho - \hat{\pi}\tanh\rho &=& 0, \label{Eq:eps_evo_visc}\\
   \hat{\tau}_\pi \frac{d \hat{\pi}}{d\rho} + \hat{\pi} + \frac{8}{3}\hat{\tau}_\pi  \hat{\pi}\tanh\rho -\frac{16}{9} \frac{\eta}{s} \frac{\hat{\epsilon}}{\hat{T}}\tanh\rho &=&0, \label{Eq:pi_evo}
\end{eqnarray}
where $\hat{\tau}_\pi$ is the shear relaxation time, proportional to shear viscosity to entropy density ratio $\eta/s$. The above equations constitute the Israel-Stewart (IS) hydrodynamic theory within Gubser flow. These equations can be solved by employing an equation of state (EoS) to close the set. We use the relativistic conformal EoS for an ideal gas of massless quarks and gluons given by
\begin{equation}\label{eos}
    \hat{\epsilon} = 3\hat{P} = 3\left[2(N_c^2 -1) + \frac{7}{2}N_c N_f \right] \frac{\pi^2}{90} \hat{T}^4,
\end{equation}
where $N_c = 3$ and $N_f=2$ are the number of colors and flavors of quarks respectively.
For a conformal fluid, $\eta \sim s$ and the relaxation time must be proportional to the inverse of temperature and we fix $\hat{\tau}_\pi = 5(\eta/s)/\hat{T}$ motivated by kinetic theory~\cite{Denicol:2010xn}.

In the absence of viscosity, Eqs.~\eqref{Eq:eps_evo_visc} and \eqref{Eq:pi_evo} reduce to a single equation for temperature, and the analytical solution of the same is given by~\cite{Gubser:2010ze,Gubser:2010ui}
\begin{eqnarray}
    \hat{T}_i (\rho) = \frac{\hat{T}(\rho_0)}{\cosh^{2/3}\rho};
\end{eqnarray}
where $\hat{T}(\rho_0) = \tau_0 T(\tau_0, r=0)$, $T(\tau_0, r=0)$ is the initial temperature and $\tau_0$ denote the initial proper time.

Now, we study the evolution of temperature and shear stress profiles of hot QCD matter obtained by numerically solving the above IS hydrodynamic equations of motion. We take the initial proper time for the hydrodynamical evolution of the system as $\tau_0=0.4$ fm, in line with QGP thermalization time estimates~\cite{Kurkela:2015qoa}. It is known that the hydrodynamic description of the system is valid when the universal scaling time $\tilde{\omega} \equiv T\tau/(4\pi\eta/s) \approx 1$~\cite{Heller:2016rtz}. We note that all the values of initial conditions used in this work confines within $\Tilde{\omega} \approx 1$. 
We choose the relevant initial conditions~\cite{Paquet:2023bdx}: $T_0=T(\tau_0,r=0) = 0.5$ GeV and $\pi_0=\pi (\tau_0, r=0)=0$ with $\tau_0 = 0.4$ fm. In Gubser coordinates, these correspond to $\hat{T}(\rho_0) = 1.02$ and $\hat{\pi}(\rho_0) =0$. The value of $\rho_0$ depends on the arbitrary length scale $q$ considered, and we
determine the same from Eq.~\eqref{Eq:rho}. We have taken $4\pi(\eta/s) = 1$ and the parameter $q=0.25, 0.35$ and 0.45 fm$^{-1}$ in our analysis, corresponding to the relevant transverse system size values. Later, we also vary viscosity as $4\pi(\eta/s) = 1$ and $2$ in our studies. In order to compare with one-dimensional boost invariant evolution of the system, we solve the IS equations in Milne coordinates for Bj\"orken flow~\cite{Denicol:2021book}
\begin{eqnarray}
    \frac{d\epsilon}{d\tau} + \frac{\epsilon + P - \pi}{\tau} &=&0, \\
   \tau_\pi \frac{d\pi}{d\tau} + \pi + \left(\frac{4}{3} + \lambda \right) \tau_\pi \frac{\pi}{\tau} - \frac{4}{3}\frac{\eta}{\tau} &=& 0,
\end{eqnarray}
where $\pi = \pi_{\eta_s}^{\eta_s}$, $\lambda = 10/21$ and $\tau_\pi = 5(\eta/s)/T$. 

We plot the temperature $\hat{T}$ and normalized shear stress $\hat{\bar{\pi}} \equiv \hat{\pi}/(\hat{\epsilon} + \hat{P})$ profiles as a function of the Gubser coordinate $\rho$, obtained by solving Eqs.~\eqref{Eq:eps_evo_visc} and \eqref{Eq:pi_evo}, for different $q$ in Fig.~\ref{fig:evolution-rho}, with $4\pi(\eta/s)=1$. It is observed that the system moves away from the initial state: $\hat{T}(\rho_0)=1.02$, $\hat{\pi}(\rho_0) = 0$ due to the initial rapid longitudinal expansion and as $\rho$ increases, the system passes through the equilibrium state $(\hat{\pi}=0)$ again. From Fig.~\ref{fig:T-rho}, we see that the temperature of the system has a maximum value at $\rho = 0$ and then decreases symmetrically with increment in $|\rho|$. With the increase in $q$, the maximum value of $\hat{T}$ observed at $\rho = 0$ descends and the temperature profile flattens. Similarly, from Fig.~\ref{fig:pi-rho}, we find that the value of $\hat{\bar{\pi}}$ first decreases below zero and reaches a minimum, then with the increase in $\rho$, $\hat{\bar{\pi}}$ also increases and passes through the initial state at $\rho = 0$. For further increment in $\rho$, $\hat{\bar{\pi}}$ keep on increasing. We observe that the minimum value of $\hat{\bar{\pi}}$ occurs at less negative value of $\rho$, when $q$ is increased.

Now, using the more familiar Milne coordinates ($\tau$ and $r$), we study the space-time dependence of temperature and shear-stress of hot QCD matter for different values of $q$. 
In subsequent analyses, we use fixed value of initial temperature and proper time $T_0=0.5$ GeV and $\tau_0=0.4$ fm. In Fig.~\ref{fig:evolution-tau}, we plot the proper-time evolution of $T$ and $\pi$ at $r=0$, with $4\pi(\eta/s)=1$. We also plot the evolution corresponding to the Bj\"orken case for comparison. From Fig.~\ref{fig:T-tau}, we observe that the temperature of the system takes longer time to cool for smaller values of $q$. This is because the scalar expansion rate within Gubser model, $\hat{\Theta} = 2\tanh\rho$ depends on the parameter $q$ and smaller values of $q$ slow down the rate of expansion which result in slower cooling of the medium. Also, we note that the temperature profile approaches the Bj\"orken case as $q \rightarrow 0$. In Fig.~\ref{fig:pi-tau}, we show the proper-time evolution of $\pi$ within Gubser model, at $r=0$. We observe that the effect of shear viscous pressure is large at early times. The impact of shear stress increases with decrease in the value of $q$. For comparison, the shear stress evolution in Bj\"orken model is also shown. Note that the peak of $\pi$ occurs almost at the same $\tau$ for the Gubser and Bj\"orken cases since relaxation time is kept fixed throughout this analysis. 

Next, we look into the complete $T$ and $\pi$ profiles by considering the transverse expansion. Fig.~\ref{fig:T-profile} shows the evolution of the system temperature in the presence of viscosity along the $r$ and $\tau$ directions, by varying the parameter $q$. For any value of $q$, it is observed that the temperature profile has a peak around the center of the fireball $i.e.,$ $r=0$, at $\tau = \tau_0$; this peak flattens over the radial coordinate with increase in the proper time. We see that the width of the peak around $r=0$ increases for smaller values of $q$; which implies that more region near $r=0$ remains at a higher temperature (region shown in red shade) for lower $q$ value, and hence the system takes longer time to cool (indicated by blue shade). 

\begin{figure}
    \centering
    \includegraphics[height=12.5cm]{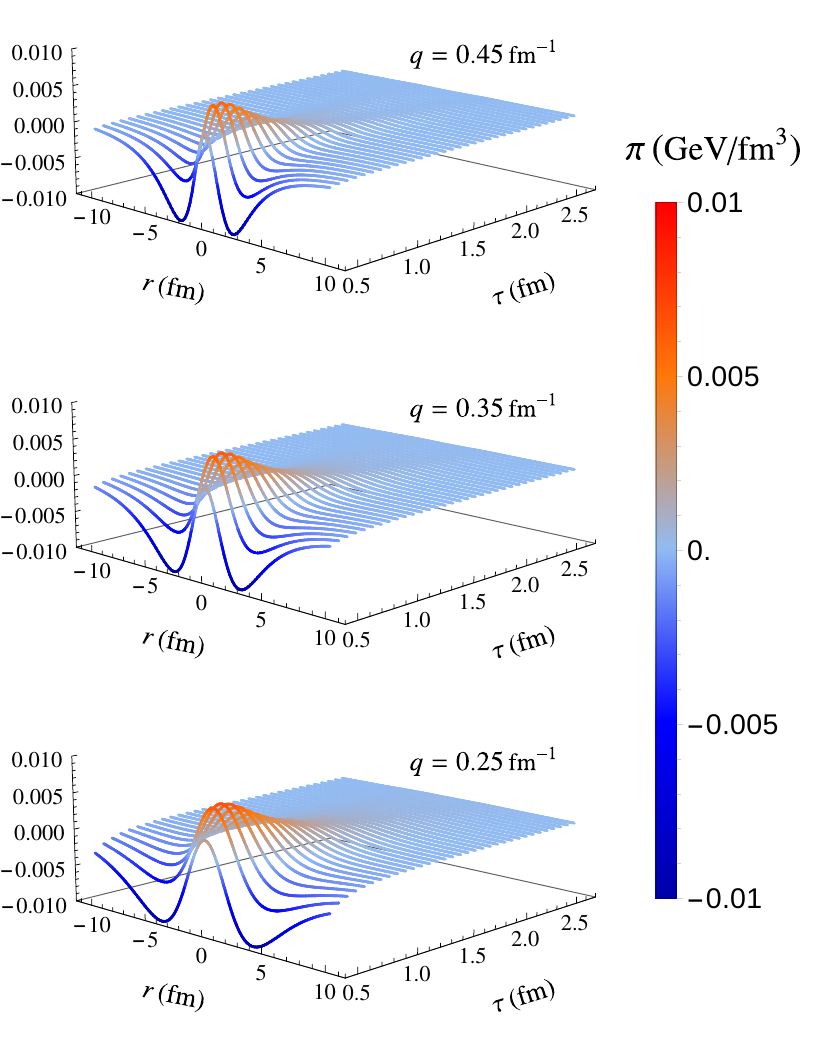}
    \caption{Shear stress evolution of QGP obtained by solving the IS hydrodynamics within Gubser flow, for different values of $q$, with $4\pi(\eta/s)=1$.}
    \label{fig:pi-profile}
\end{figure}

Similarly, in Fig.~\ref{fig:pi-profile}, we depict the shear stress profile $\pi$ as a function of $r$ and $\tau$, for different values of $q$. We note that the effect of shear pressure is more visible around the center of the system and at early proper times. At initial times, $\pi$ profile begins from a positive peak value at $r=0$ and approaches a negative peak with increment in $r$ and then increases finally to zero for large $r$ values. As $\tau$ increases, in the region near $r=0$ (shown in red), $\pi$ first increases and reaches a positive maximum value, then starts to decrease and finally approaches zero. Whereas, as we move away from the centre (i.e. region depicted in blue), as $\tau$ evolves, $\pi$ monotonously increases from the negative values to zero. It can be seen that, with sufficient increment in $\tau$, the negative peaks observed at either side of $r=0$ decreases and flatten over $r$ to the value $\pi =0$. Further, when $q$ is increased from 0.25 to 0.45 fm$^{-1}$, the width of initial $\pi$ profile decreases. Also we observe that, for large $q$, the shear stress evolution happens faster.
\section{Thermal dilepton spectra within Gubser flow} \label{Sec:dilepton-spectra}

In this section, we obtain the thermal dilepton spectra from heavy-ion collisions by employing the Gubser symmetry. Dileptons are emitted from all the stages of collisions through various mechanisms and probe the entire temperature history of collisions. Here, we focus on the dominant mechanisms of thermal dilepton production from QGP and hadronic phases. The major contribution of thermal dilepton emission from QGP medium comes from the $q\bar{q}$-annihilation process, $q\bar{q} \rightarrow \gamma^\ast \rightarrow l^{+}l^{-}$ and in the hadronic phase, $\pi^{+}\pi^{-}$-annihilation, $\pi^{+}\pi^{-} \rightarrow \rho^{0} \rightarrow l^{+}l^{-}$ contributes dominantly to the spectra. From relativistic kinetic theory, the rate of dilepton production for these processes is given by~\cite{Vogt:2007zz}
\begin{eqnarray} \label{Eq:dil_rate}
 \frac{dN}{d^4x d^4p} &=& \int \frac{d^3 {\bf p_1}}{(2\pi)^3} \frac{d^3{\bf p_2}}{(2\pi)^3}\,
 \frac{M^2 g^2 \sigma(M^2)}{2 E_1 E_2}\nonumber\\
 && \times f_q({\bf p_1}) f_q({\bf p_2}) \delta^4(p-p_1- p_2).
\end{eqnarray}

\begin{figure}[b]
    \centering
    \includegraphics[width=\linewidth]{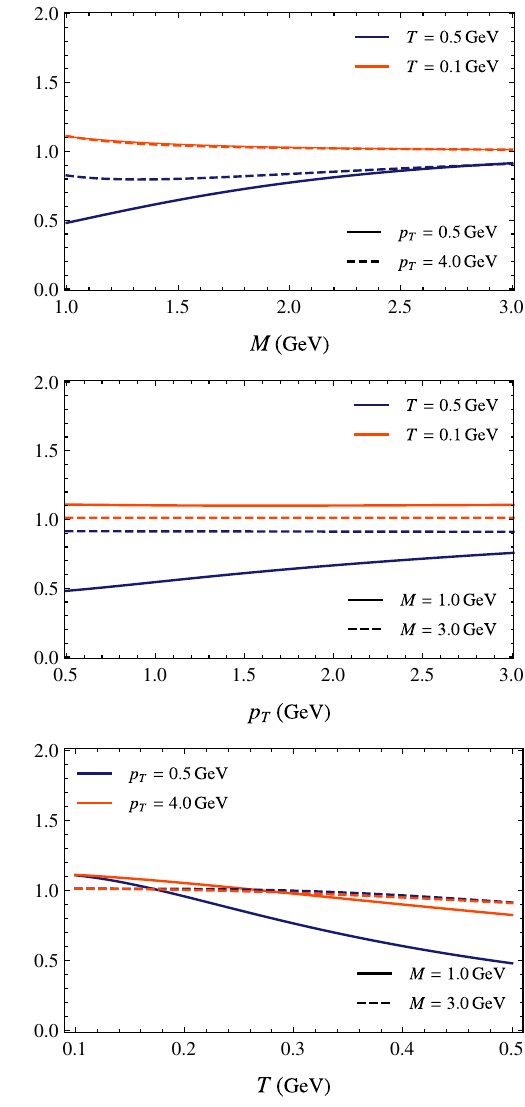}
    \caption{Ratio of thermal dilepton rate from hadronic medium considering the massless pion approximation ($m_\pi =0$) to that calculated with $m_\pi=0.139$ GeV. 
}
    \label{fig:rate-ratio}
\end{figure}

In the above expression, $p_{1,2}=(E_{1,2},{\bf p_{1,2}})$ denote the four-momenta of incoming particles having masses $m_{1,2}$. Four-momentum of the lepton pair is $p = (E=E_1 + E_2, {\bf p} = {\bf p_{1}} + {\bf p_{2}})$. Here, $\sigma(M^2)$ represent the cross-section for the process in the Born approximation, where $M^2 = (E_1 + E_2)^2 - ({\bf p_{1}} + {\bf p_{2}})^2$ denotes the invariant mass of the virtual photon, and $g$ is the degeneracy factor. Taking $N_f = 2$ and $N_c = 3$, we have $M^2 g^2 \sigma_q(M^2) = 80\pi/9 \alpha_e^2$ for the $q\bar{q}$-annihilation process in QGP phase, with $\alpha_e$ being the electromagnetic coupling constant. Also, we have $M^2 g^2 \sigma_\pi(M^2) = (4\pi/3)\alpha_e^2 |F_\pi(M^2)|^2$~\cite{Vogt:2007zz} for $\pi^{+}\pi^{-}$-annihilation in the hadronic phase. The term $|F_\pi(M^2)|^2 = m_\rho^4/[(m_\rho^2 - M^2)^2 + m_\rho^2 \Gamma_\rho^2]$ is the electromagnetic pion form factor with $m_\rho = 775$ MeV and $\Gamma_\rho =149$ MeV being the mass and decay width of $\rho(770)$ meson respectively~\cite{Song:2010fk}. Further, $f_i(\textbf{p}_{1,2})$ are the phase-space distribution functions of interacting particles, where $i\equiv(q,\pi)$.

Presence of viscosity affects the dilepton production in two ways: first, through the hydrodynamic expansion of hot QCD medium and second through the distribution function. Viscous modifications to the phase-space distribution functions can be taken as $f_i = f^0_i + \delta f_i$, where $\delta f_i$ denotes the correction due to viscosity and $f^0_i$ is the equilibrium distribution function. Here, we use the $\delta f_i$ upto first-order obtained from the Chapman-Enskog like method~\cite{Bhalerao:2013pza}
\begin{equation}\label{CE_deltaf}
    \delta f_i = \frac{f^0_i\,\beta}{2 \beta_\pi (u \cdot p)} p^\mu p^\nu \pi_{\mu\nu},
\end{equation}
where $\beta = 1/T$ and $\beta_\pi = (\epsilon + P)/5$. By assuming the Maxwell-Boltzmann form for the equilibrium part and substituting the viscous modified distribution function in Eq.~\eqref{Eq:dil_rate}, the expression for dilepton rate (keeping the terms upto second order in momenta) can be written as the sum of ideal and shear viscous contributions
\begin{eqnarray}
   \frac{dN}{d^4x d^4p} = \frac{dN^0}{d^4x d^4p} + \frac{dN^\pi}{d^4x d^4p}.
\end{eqnarray}
Now, the ideal contribution to the dilepton rate is obtained as~\cite{Vogt:2007zz} 
\begin{eqnarray} \label{Eq:rate_ideal}
\frac{dN^0}{d^4x d^4p} &=& 
\frac{1}{2}\sum_{i=q,\pi}\mathcal{R}_i e^{-u\cdot p/T},
\end{eqnarray}
where $\mathcal{R}_i =M^2g^2\sigma_i(M^2)/(2\pi)^5$. We obtain the shear viscous part of dilepton rate as~\cite{Naik:2021yph,Naik:2021yue}
\begin{eqnarray} \label{Eq:rate_shear}
    \frac{dN^\pi}{d^4x d^4p} &=&  \frac{\beta}{\beta_\pi} \frac{p^\mu p^\nu \pi_{\mu\nu}}{2\mathcal{E}^{5}} \Bigg(  \mathcal{E}\left[\mathcal{E}^2-\frac{3}{2}M^2\right] (u\cdot p) \nonumber\\ &&+\frac{3}{4}M^4 \ln\left[\frac{(u\cdot p)+\mathcal{E}}{(u\cdot p)-\mathcal{E}}\right]\Bigg)\frac{dN^{0}}{d^4x d^4p};
    \end{eqnarray}
    where, $\mathcal{E}^2= (u\cdot p)^2-M^2.$

The above rate expressions (Eqs.~\eqref{Eq:rate_ideal} and \eqref{Eq:rate_shear}) are integrated over the space-time history of heavy-ion collisions along with the temperature and shear stress tensor profiles of hot quark-gluon matter to obtain the thermal dilepton yield. 

We now obtain the quantities $(u \cdot p)$ and $p^\mu p^\nu \pi_{\mu\nu}$ appearing in the dilepton rate expressions under Gubser flow. The four-momentum of the dileptons can be parameterized as $p^\mu = \Big(m_T \cosh(y-\eta_s), p_T \cos(\phi_p - \phi), p_T \sin(\phi_p - \phi)/r, m_T \sinh(y-\eta_s)/\tau \Big)$; where $m_T =\sqrt{p_T^2 + M^2}$ is the transverse mass of the dilepton, with $p_T$ and $M$ are its transverse momentum and invariant mass respectively. Also, $\phi_p$ is the phase angle and $y$ denotes the rapidity of the particle. Using the definitions of $\hat{u}^\mu$ and $\hat{\pi}^{\mu\nu}$ within Gubser flow, along with $p^\mu$, we obtain the factors appearing in the rate expressions as: 
\begin{eqnarray}
u\cdot p &=& u_\tau m_T \cosh(y-\eta_s) - u_r p_T \cos(\phi_p - \phi),\,\,\,\,\quad \label{Eq:udotp} \\
    p^\mu p^\nu \pi_{\mu\nu} &=& 
 m_T^2 \left( \cosh^2(y - \eta_s) \,\pi_{\tau\tau} + \frac{\sinh^2(y - \eta_s)}{\tau^2}\,\pi_{\eta_s \eta_s}\right)\nonumber \\ 
 &+& p_T^2 \left(  \cos^2(\phi_p - \phi) \pi_{rr} 
+ \frac{\sin^2(\phi_p - \phi)}{r^2} \pi_{\phi\phi}\right); \label{Eq:psquare}
\end{eqnarray}
where 
\begin{eqnarray}
 u_\tau  &=& \frac{\partial \rho}{\partial \tau} u_\rho = \frac{1}{\sqrt{1-v_{r}(\tau,r)^2}},  \\
 u_r &=& \frac{\partial \rho}{\partial r} u_\rho = \frac{-v_{r}(\tau,r)}{\sqrt{1-v_{r}(\tau,r)^2}}; \label{flow_milne}
\end{eqnarray} 
with
\begin{eqnarray}
 v_{r}(\tau,r) &=& \frac{2q^2\tau r}{1+(q\tau)^2 + (qr)^2}; \nonumber
\end{eqnarray}
and 
\begin{eqnarray}
\pi_{\tau\tau} &=& -\frac{\hat{\pi}}{2\tau^2}\left( \frac{\partial \theta}{\partial \tau}\right)^2\cosh^2\rho, \\
\pi_{rr} &=&-\frac{\hat{\pi}}{2\tau^2}\left( \frac{\partial \theta}{\partial r}\right)^2\cosh^2\rho, \\
\pi_{\phi\phi} &=&-\frac{\hat{\pi}}{2\tau^2}\cosh^2\rho \sin^2\theta;
\end{eqnarray}
with 
\begin{equation}
\pi_{\eta_s \eta_s} = \frac{\hat{\pi}}{\tau^2}.    
\end{equation}
Noting the four-dimensional volume element to be $d^4 x = \tau d\tau\,rdr\,d\phi\,d\eta_s$, the dilepton yield from heavy-ion collisions within the Gubser flow can be now calculated as
\begin{eqnarray} \label{Eq:dil-yield-1}
    \frac{dN}{dM p_Tdp_T dy} 
    &=& M\int_0^{2\pi} d\phi_p \int_{\tau_0}^{\infty} \tau\,d\tau \int_0^\infty r\,dr \int_0^{2\pi}d\phi \nonumber\\
    &&\times\int_{-\infty}^{\infty}d\eta_s \left[\frac{dN}{d^4x d^4p} \right] \Theta({\scriptstyle T > T_{\textrm{min}}}),
\end{eqnarray}
where the Heaviside step function ($\Theta$) restricts the evolution of temperature till $T_\textrm{min}$. The ideal contribution to the thermal dilepton yield can be simplified by performing the integrals over $\eta_s$ and $\phi$ analytically using the integral representation of modified Bessel functions of first ($K_n (z)$) and second ($I_n (z)$) kinds respectively given in Appendix.~\ref{Bessel}. The expression for ideal part of the dilepton yield can be now written as
\begin{eqnarray} \label{Eq:id-yield}
     \frac{dN^{0}}{dM p_T dp_T dy} 
    &=& 2\pi M \sum_{i=q,\pi}\mathcal{R}_i \int_0^{2\pi} d\phi_p \int_{\tau_0}^{\tau_f} \tau\,d\tau\ \int_0^\infty r\,dr \nonumber\\
    &\times&  K_0\big[\frac{u_\tau m_T}{T} \big] I_0\big[\frac{-u_r p_T}{T}\big]\,\,\Theta({\scriptstyle T > T_{\textrm{min}}}).
\end{eqnarray}
Similarly, the $p_T$ integrated differential yield $dN/dMdy$
is obtained as 
\begin{eqnarray} \label{Eq:dil-yield-2}
    \frac{dN}{dMdy} &=& \int_{p_T^{min}}^{p_T^{max}} p_T\, dp_T \left[ \frac{dN}{dM p_Tdp_T dy} \right]. 
\end{eqnarray}
\begin{figure*}
    \centering
    \subfigure
[]{\includegraphics[width=8.5cm]{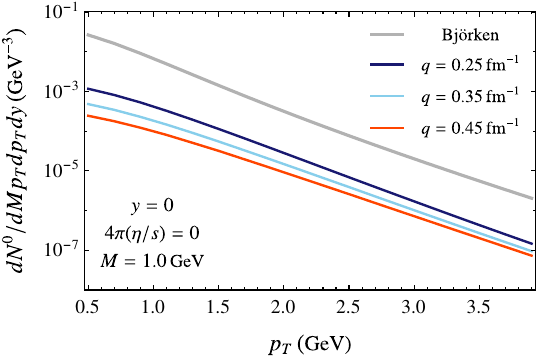} \label{fig:dil-id}}\quad
\subfigure
[]{\includegraphics[width=8.5cm]{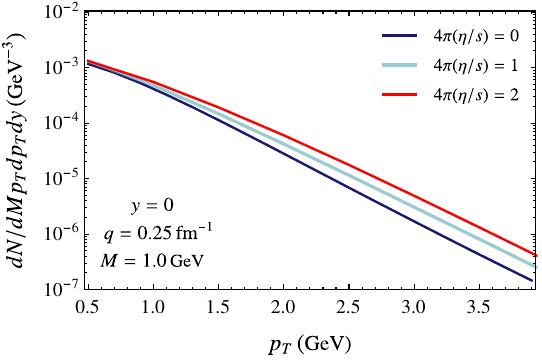} \label{fig:dil-visc-1}}
    \caption{(a) Thermal dilepton yield for relativistic ideal hydrodynamics by varying the arbitrary parameter $q$, with $M=1.0$ GeV. The yield corresponding to the ideal Bj\"orken scenario $(q\to 0)$ is also shown for comparison. (b) Thermal dilepton yield from IS hydrodynamics within Gubser flow by varying the viscosity, for $q=0.25$ fm$^{-1}$. }
\end{figure*}

\begin{figure}
\centering
\includegraphics[width=9cm]{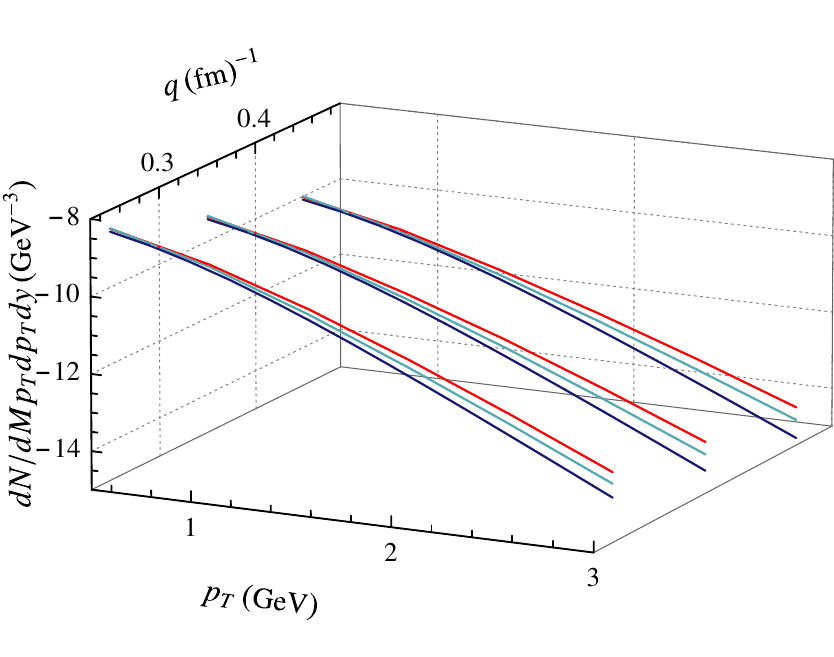}
\caption{Thermal dilepton yield from QGP and hadronic matter by varying the parameter $q$, for $M=1.5$ GeV. The solid dark blue, light blue and red curves denote the yields corresponding to $4\pi(\eta/s)=0, 1$ and 2 respectively. }\label{fig:dil-3d}
\end{figure}

By employing the temperature and viscous $(\pi)$ profiles of the hot fireball obtained in the previous section, we numerically evaluate Eqs.~\eqref{Eq:dil-yield-1} and \eqref{Eq:dil-yield-2} to obtain the dilepton spectra.
\par 
Further, for comparison with Chapman-Enskog like distribution function, we calculate the dilepton yield using the 14-moment Grad's non-equilibrium correction~\cite{Dusling:2007gi}
\begin{eqnarray} \label{grads_deltaf}
\delta f_G = \frac{f^0 }{2sT^3}p^\mu p^\nu \pi_{\mu\nu}.
\end{eqnarray}
The shear viscous contribution to the dilepton rate due to the above correction is obtained as~\cite{Dusling:2008xj}
\begin{eqnarray}
    \frac{dN^\pi_G}{d^4x d^4p} = \frac{2}{3}\left(\frac{p^\mu p^\nu}{2sT^3}\pi_{\mu\nu}\right)\frac{dN^0}{d^4x d^4p},
\end{eqnarray}
where $s = (\epsilon + P)/T$ is the entropy density. The ideal part is given by Eq.~\eqref{Eq:rate_ideal}. One can obtain the corresponding dilepton yield within Gubser geometry in similar manner by making use of the expressions discussed before.  
\par 
Before we proceed towards numerical evaluation of the dilepton yields within Gubser flow in the next session, we would like to comment on the validity of the assumption $m_\pi =0$ in the dilepton rate calculation from hadronic medium. We plot the ratio of dilepton rate (ideal case) from hadronic phase calculated by considering the massless pion approximation to that obtained with non-zero mass $m_\pi=0.139$ GeV, by varying $M$, $p_T$ and $T$ (keeping $p_z=0$) in Fig.~\ref{fig:rate-ratio}. While calculating the rate in the massive case, we have also used the Bose-Einstien distribution function for pions (See Appendix \ref{dilepton_massive} for the rate calculation). We see that the approximation works better for all high values of $M$ regardless of the $p_T$ and $T$ values considered (error remains less than 10\%). At lower values of $M$, low $p_T$ and high $T$ result in more error. One can see that errors are well within 10\%, except for the lower values of $M$ and $p_T$ ($\sim 50\%$ error). Therefore considering $m_\pi =0$ is a reasonable approximation in our calculations and since Gubser flow is valid only with the conformal EoS, it is only consistent to use massless limit for pions in our study.

\section{Results and discussion} \label{Sec:results}
\begin{figure*}
\centering
\subfigure
[]{\includegraphics[width=8.5cm]{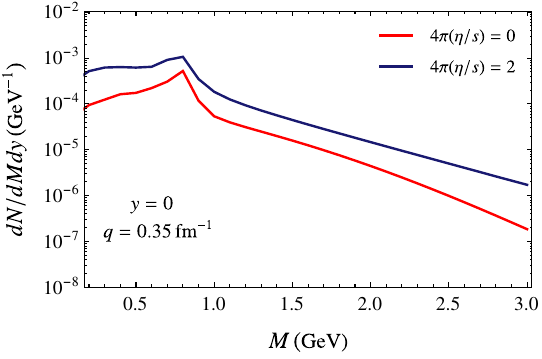} \label{fig:dNdMdy}}\quad
\subfigure[]{\includegraphics[width=8.5cm]{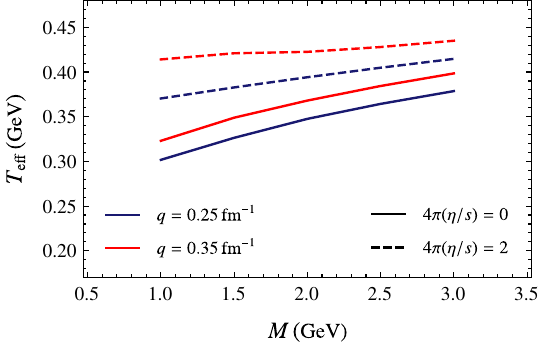}\label{fig:Teff}}
\caption{(a) Dilepton invariant mass spectra from both viscous and inviscid expanding hot QCD medium, by fixing $q=0.35$ fm$^{-1}$. (b) Effective temperature of the system as a function of invariant mass of dilepton. Solid lines denote the ideal $T_{\textrm{eff}}$ calculated from ideal spectra.}
\end{figure*}

In this section, we analyze the thermal dilepton spectra from heavy-ion collisions obtained by employing the solutions of Israel-Stewart hydrodynamics within Gubser flow. As already specified in Sec.~\ref{Sec:Israel-Stewart}, we adopt the initial conditions $T(\tau_0,r=0) = 0.5$ GeV, $\pi (\tau_0, r=0)=0$ at $\tau_0 = 0.4$ fm and
we choose three different values of arbitrary energy scale, $q=0.25, 0.35,$ and 0.45 fm$^{-1}$ in our analysis. We make use of the temperature and shear stress profiles obtained in Sec.~\ref{Sec:Israel-Stewart} by numerically solving Eqs.~\eqref{Eq:eps_evo_visc} and \eqref{Eq:pi_evo}. Since we consider dileptons from both QGP and hadronic sources, we evolve the temperature \textcolor{blue}{up to} a minimum value, $T(\tau_f, r) = T_{\text{min}}$, where $\tau_f$ is the final proper time. Note that, here the value of $\tau_f$ depends on the coordinate $r$ and hence 
matter at different radial distances from the center of fireball cool to 
$T_{\text{min}}$ at different proper times. 
The Heaviside Theta function $\Theta({\scriptstyle T > T_{\textrm{min}}})$ in the dilepton yield expression takes this into consideration in our numerical analysis. 
Following Ref.~\cite{Paquet:2023bdx}, we take $T_{\text{min}}=0.1$ GeV.
Also note that the yields are plotted for midrapidity region of dileptons $i.e.,$ $y=0$.  

We first study the effect of variation of parameter $q$ on the dilepton spectra by plotting the ideal dilepton yield in Fig.~\ref{fig:dil-id}, for the invariant mass $M = 1.0$ GeV. The yield obtained within the Bj\"orken model, for the ideal temperature profile, $T(\tau) = T_0 (\tau_0/\tau)^{1/3}$ is also shown for comparison. It is observed that there is an overall suppression in the spectra obtained for the Gubser flow for any $q$, compared to the Bj\"orken scenario. This is because, within the one-dimensional Bj\"orken model, temperature of QGP takes longer time to cool, in contrast to the Gubser case; as a result, the number of dileptons produced will be more. Also, it is seen that there is an overall decrement in the spectra with an increment in $q$ value. This decrement is profound in the low $p_T$ regime of the spectra. 
We note that, the parameter $q$ is inversely related to the transverse radius/size of the fireball, $L = 1/q$. Therefore, an increment in $q$ implies a decrease in the transverse size resulting in less production of lepton pairs. Further, we found that the spectra tend to approach the Bj\"orken case when $q \rightarrow 0$, as expected.
\par 
In order to understand whether the dilepton spectra from Gubser will approach to that of Bj\"orken, we vary $M$ and $p_T$ to high values and check their convergence. We keep the initial time ($\tau_0=0.4$ fm) and temperature ($T_0=0.5$ GeV) along with the final evolution temperature same for both the cases, with $q=0.45$ fm$^{-1}$. We note that for the ideal Bj\"orken case, it takes $\sim 50$ fm to reach the $T_f=0.1$ GeV, whereas for the Gubser it is only $\sim 3.7$ fm at the center of the fireball ($r=0$). It was found that even with very high values of $M$ ($\sim$ 4 GeV) and $p_T$ ($\sim$ 10 GeV), the Gubser spectra will not converge to that of Bj\"orken. Our studies indicate that the convergence of dilepton spectra within Gubser and Bj\"orken models happens only with the condition $q\rightarrow0$.

\par 
Next, we analyze the effect of viscosity on the dilepton yield  obtained within Gubser solutions of IS hydrodynamics in Fig.~\ref{fig:dil-visc-1}, by varying the viscosity. We fix the $q$ value to be $0.25$ fm$^{-1}$ for this analysis. The solid dark blue curve denotes the ideal dilepton yield. We consider the viscosities $4\pi(\eta/s)=1, 2$ and are represented using solid light blue, red curves respectively. The presence of viscosity enhances the dilepton yield, since incorporating the viscous corrections to the dilepton production rates result in positive contribution to the spectra. Also, the viscous effects in the medium slows down the expansion of the fireball and thereby increases the yield~\cite{Bhatt:2010cy,Bhatt:2011kx}. In Fig.~\ref{fig:dil-3d}, we show the effect of variation of the parameter $q$ on the viscous dilepton spectra for $M=1.5$ GeV in a three-dimensional plot. As mentioned before, we observe an overall decrement in the dilepton spectra when $q$ is increased. 

In Fig.~\ref{fig:dNdMdy}, we plot the $p_T$-integrated dilepton yield from both viscous and ideal hot QCD medium as a function of dilepton invariant mass $M$. For this analysis, we have fixed $q=0.35$ fm$^{-1}$ and for the viscous case, the viscosity is taken to be $4\pi(\eta/s) = 2$. The integration over $p_T$ is carried out in the range: $1\leq p_T\,(\text{GeV})\leq 20$~\cite{Ryblewski:2015hea}. As expected, we find an overall enhancement in the invariant mass spectra due to the presence of viscosity and this increment is more profound for higher invariant masses. 
We note that the peak observed at $M=0.77$ GeV is because of the dilepton production from $\rho(770)$ decay. 

Next, in order to have a complete analysis of the thermal dileptons within viscous Gubser flow, we determine the effective slope parameters of transverse momentum spectra. We follow the method outlined in Ref.~\cite{Dusling:2008xj} and calculate the effective temperature $T_{\textrm{eff}}$ from the dilepton spectra as a function of invariant mass. We find $T_{\textrm{eff}}$ by fitting the spectra at a fixed $M$ with the expression:
\begin{eqnarray}
       \frac{dN}{dM^2 m_Tdm_T dy}  \propto \textrm{exp}\left( -m_T/T_\textrm{eff}\right).
\end{eqnarray}
We choose the range $1.1 \leq p_T (\textrm{GeV})\leq 2.1$ for this fit. In Fig.~\ref{fig:Teff}, we plot the $T_\text{eff}$ determined for ideal and viscous dilepton spectra as a function of $M$, for different $q$ values. There is an enhancement in the $T_\text{eff}$ curve with increase in $M$, implying that the viscous corrections increase with $M$.
We observe that the $T_\text{eff}$ determined from transverse momentum spectra within viscous Gubser hydrodynamics is larger compared to that evaluated within ideal Gubser solution. Further, the $T_\textrm{eff}$ values are high for large value of $q$, since the effective temperature is determined from the inverse slope of the spectra. 

We must note that in the analysis so far, we have fixed the initial temperature $T(\tau_0, r=0) \equiv T_0$ while studying the variation of the $q$ parameter. This results in different initial energy density profiles for the system with peak value at $r=0$ remaining the same but with narrowed radial spread, as we increase the $q$ value (smaller system size). This corresponds to higher initial energy density for smaller systems, even though its total energy is lesser compared to larger systems. And as we saw, thermal spectra with such initial conditions give rise to higher effective temperature for a smaller system. We now fix the initial total energy ($E_\text{tot}$) of the system and vary $q$ in the analysis, thus covering both the initial energy and system size effects. We first obtain the value of $E_\textrm{tot}$ for $T_0=0.5$ GeV and $q=0.25$ fm$^{-1}$. Then fixing thus obtained $E_\text{tot}$ value, we determine $T_0$ corresponding to $q=0.35$ fm$^{-1}$. This results in higher temperature or initial energy density value at $r=0$ for larger $q$, i.e., smaller system. Also, the initial energy density profile gets narrowed for large $q$, since $E_\textrm{tot}$ is kept constant. In this scenario too, we find that smaller system results in a higher effective temperature. While fixing $T_0$, the temperature profile becomes increasingly localized around the center with increment in $q$. This leads to a larger fraction of the total emission of dileptons originating from hotter regions. In the second case, maintaining the same total energy in a smaller volume necessitates a higher initial temperature, again enhancing the emission of high-momentum particles. Although this trend may appear counter-intuitive when compared to phenomenological expectations that associate larger systems with higher temperatures, it arises naturally from the spatial structure of the energy density in conformal Gubser hydrodynamics. These observations highlight the fact that effective temperatures extracted from thermal spectra are sensitive not only to the magnitude of the initial temperature or energy, but also to their spatial distribution.

\par
\begin{figure}
    \centering
    \includegraphics[width=\linewidth]{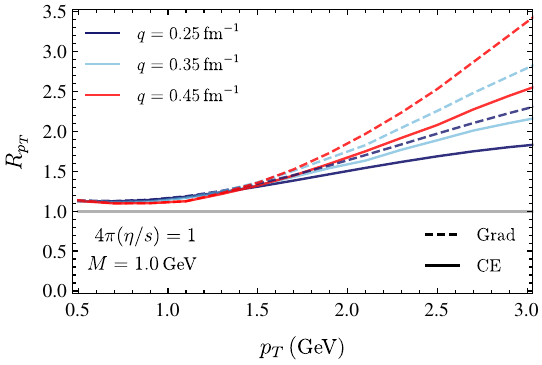}
    \caption{Ratios of particle yields for  Chapman-Enskog like (Eq.~\eqref{CE_deltaf}) and Grad's (Eq.~\eqref{grads_deltaf}) distribution functions for different $q$ values.}
    \label{fig:ratio}
\end{figure}

 
Finally, in Fig.~\ref{fig:ratio}, we compare the dilepton spectra obtained for Chapman-Enskog like viscous correction (Eq.~\eqref{CE_deltaf}) with that obtained using 14-moment Grad's method (Eq.~\eqref{grads_deltaf}), for various $q$ values. We fix $M = 1.0$ GeV and $4 \pi (\eta/s) =1$ for this comparison with fixing the central temperature $T(\tau_0, r=0) = 0.5$ GeV, as before. We plot the ratio defined by
\begin{eqnarray}
    R_{p_T} = \left[\frac{dN}{dM p_T dp_T dy} \right]/\left[\frac{dN^0}{dM p_T dp_T dy}\right],
\end{eqnarray}
which gives the non-equilibrium corrections to the ideal dilepton yield (Eq.~\eqref{Eq:id-yield}). It can be seen that the viscous corrections grow with $p_T$ for any value of $q$. This can be understood from the term $p^\mu p^\nu \pi_{\mu\nu}$ (Eq.~\eqref{Eq:psquare}) in the $\delta f$, which is proportional to $p_T^2$. It must be noted that the Grad's viscous correction is larger compared to the CE-like correction for any value of $p_T$ and $q$. This will result in higher particle spectra while employing $\delta f_G$~\cite{Bhalerao:2013pza}. Also, the difference between both the viscous corrections is found to increase for larger $p_T$ values. Further, we see that the strength of viscous corrections increase with increment in the value of $q$. We conclude from this analysis that the CE like $\delta f$ is more suitable to study particle production from heavy-ion collisions, as indicated by the earlier study using 1-D Bj\"orken flow~\cite{Bhalerao:2013pza}.

\section{Conclusion} \label{Sec:conclusion}

We have employed the solutions of second-order Israel-Stewart hydrodynamics within Gubser flow to investigate the thermal dilepton production from heavy-ion collisions. We calculated the dilepton rate in the presence of first-order Chapman-Enskog (CE) like viscous correction. The dilepton yield is calculated within Gubser model, which considers the transverse dynamics along with the longitudinal boost-invariance. The inclusion of radial dependence in the dilepton yield resulted in an overall decrement of the dilepton spectra compared to that obtained within 1-D Bj\"orken expansion. The radial expansion of QGP has profound impact on dilepton yields and we found that the dilepton spectra changes appreciably with variation in the value of arbitrary parameter $q$. We have also explored the effect of viscosity on the dilepton yields.

Further, we determined the effective temperature of QGP medium ($T_{\textrm{eff}}$) from the inverse slope of the transverse mass spectra obtained within Gubser flow.  
Our study reveals that in Gubser flow, smaller systems (larger $q$) yield higher effective temperatures, regardless of whether the initial condition fixes the temperature $T(\tau_0, r=0)$ or total energy. This emanates from the increasingly localized energy density profiles in smaller systems, which enhance emission of particles from hotter regions. Our results emphasize that spatial geometry plays a critical role in shaping thermal spectra and effective temperature observables.
\par 
Finally, we have compared the strength of Chapman-Enskog (CE) like viscous correction to that of 14-moment Grad's and found that the strength of Grad's correction to the ideal dilepton yield is significantly high, especially at high $p_T$. Therefore, the CE like viscous correction is preferred over Grad's correction for studying the particle spectra. Furthermore, it will be interesting to employ the Gubser model to study the electromagnetic signals by considering CE like viscous corrections up to second-order. We leave this analysis for future study.

\section*{Acknowledgments}
\par 
Authors would like to thank the anonymous referee, whose suggestions have improved the quality of the manuscript.
L. J. N. acknowledges the Department of Science and Technology, Government of India for the
INSPIRE Fellowship.

\appendix

\section{Integral representation of modified Bessel functions} \label{Bessel}

The expression for ideal contribution to dilepton yield Eq.~\eqref{Eq:id-yield} is obtained by using the following integral representations of modified Bessel functions of first and second kind respectively:
\begin{eqnarray}
I_n(z) &=& \frac{1}{\sqrt{\pi}\Gamma(n+1/2)} (z/2)^n \int_0^\pi  e^{-z\cos(t)} \sin^{2n}(t) dt , \nonumber\\
K_n(z) &=& \frac{\sqrt{\pi}}{\Gamma(n+1/2)} (z/2)^n \int_0^\infty e^{-z\cosh(t)} \sinh^{2n}(t) dt, \nonumber\\
\end{eqnarray}
where $\Gamma(n)$ denotes the Gamma function.

\section{Thermal dilepton production rate in hadronic medium considering massive pions} \label{dilepton_massive}
Here we derive the expression for thermal dilepton production rate in the hadronic medium by considering masses of incoming pions $(m_{\pi})$. We start with the expression
\begin{eqnarray} \label{Eq:dil_rate_1}
 \frac{dN}{d^4x d^4p} &=& \int \frac{d^3 {\bf p_1}}{(2\pi)^3} \frac{d^3{\bf p_2}}{(2\pi)^3} v_{\text{rel}} \sigma_\pi(M^2) \nonumber\\
&&\times f({\bf p_1}) f({\bf p_2}) \delta^4(p-p_1- p_2),
\end{eqnarray}
where $v_\text{rel} = \sqrt{M^2(M^2 - 4m_{\pi}^2)}/(2E_1E_2)$ is the relative velocity and the cross-section is given by
\begin{equation}
    \sigma_\pi(M^2) = \frac{4\pi\alpha^2}{3}\frac{|F_\pi(M^2)|^2}{M^2}\sqrt{1-\frac{4m_\pi^2}{M^2}}.
\end{equation}
$f({\bf p})$ is the Bose-Einstein distribution function.
Integrating Eq.~\eqref{Eq:dil_rate_1} over ${\bf p_2}$ and expressing the momenta ${\bf p}$ and ${\bf p_1}$ in terms of spherical polar coordinates, we obtain
\begin{eqnarray}
 \frac{dN}{d^4x d^4p} &=&  \int_{{p_1}_{\text min}}^{{p_1}_{\text max}} \frac{d|{\bf p_1}|}{(2\pi)^5} |{\bf p_1}|  \int_{-1}^{+1} d(\cos\theta) \nonumber\\
 &&\times \frac{\sqrt{M^2(M^2 - 4m_{\pi}^2)}}{2E_1 |{\bf p}|} \sigma_\pi(M^2)  \nonumber\\
  &&\times f({\bf p_1}) f({\bf p -p_1}) \delta(\cos\theta - \cos\theta_0),
\end{eqnarray}
where $\cos\theta_0 = (2E\sqrt{ |{\bf p_1}|^2 + m_{\pi}^2} - M^2)/2pp_1$. Now, performing the integral over the $\delta$ function, we get
\begin{eqnarray} \label{Eq:dil_rate_3}
  \frac{dN}{d^4x d^4p} &=& \int_{{p_1}_{\text min}}^{{p_1}_{\text max}}\frac{d|{\bf p_1}|}{(2\pi)^5} |{\bf p_1}|    \frac{\sqrt{M^2(M^2 - 4m_{\pi}^2)}}{2E_1 |{\bf p}|} \sigma_\pi(M^2) \nonumber\\ 
  &&\times  f({\bf p_1}) f({\bf p -p_1})\Bigg|_{\cos\theta_0}.
\end{eqnarray}
The remaining $|{\bf p}_1|$ integral is evaluated numerically to obtain the dilepton rate with limits of integration fixed from $-1 \leq \cos\theta_0 \leq +1$. We use the above rate expression to compare with that obtained in the massless pion approximation in Sec.~\ref{Sec:dilepton-spectra}. Note that in the limit $m_\pi=0$ and while considering the Maxwell-Boltzmann statistics, Equation.~\eqref{Eq:dil_rate_3} reduces to the rate for pion annihilation in Eq.~\eqref{Eq:rate_ideal}. 

\bibliography{dil-Gb}
\end{document}